# Fermi Surface Topology and Magneto-transport Properties of Superconducting $Pd_3Bi_2Se_2$

Ramakanta Chapai,[1,*] Gordon Peterson,[1] M. P. Smylie,[1,2] Xinglong Chen,[1] J. S. Jiang,[1] David Graf,[3] J. F. Mitchell,[1] and Ulrich Welp[1]

[1]*Materials Science Division, Argonne National Laboratory, Lemont, IL 60439, USA*
[2]*Department of Physics and Astronomy, Hofstra University, Hempstead, NY 11549, USA*
[3]*National High Magnetic Field Laboratory, Tallahassee, FL 32310, USA*

$Pd_3Bi_2Se_2$ is a rare realization of a superconducting metal with a non-zero $Z_2$ topological invariant. We report the growth of high-quality single crystals of layered $Pd_3Bi_2Se_2$ with a superconducting transition at $T_c \approx 0.80$ K and upper critical fields of ~10 mT and ~5 mT for the in-plane and out-of-plane directions, respectively. Our density functional theory (DFT) calculations reveal three pairs of doubly degenerate bands crossing the Fermi level, all displaying clear three-dimensional dispersion consistent with the overall low electronic anisotropy (<2). The multiband electronic nature of $Pd_3Bi_2Se_2$ is evident in magneto-transport measurements, yielding a sign-changing Hall resistivity at low temperatures. The magnetoresistance is non-saturating and follows Kohler's scaling rule. We interpret the magneto-transport data in terms of open orbits that are revealed in the DFT-calculated Fermi surface. de Haas-van Alphen (dHvA) oscillation measurements using torque magnetometry on single crystals yield four frequencies for out-of-plane fields: $F_\alpha = (150 \pm 26)$ T, $F_\beta = (293 \pm 10)$ T, $F_\gamma = (375 \pm 20)$ T and $F_\eta = (1017 \pm 12)$ T, with the low frequency dominating the spectrum. Through the measurement of angular dependent dHvA oscillations and DFT calculations, we identify the $F_\alpha$ frequency with an approximately ellipsoidal electron pocket centered on the $L_2$ point of the Brillouin zone. Lifshitz-Kosevich analysis of the dHvA oscillations reveals a small cyclotron effective mass $m^* = (0.11 \pm 0.02)m_0$ and a nontrivial Berry phase for the dominant orbit. The presence of nontrivial topology in a bulk superconductor positions $Pd_3Bi_2Se_2$ as a potential candidate for exploring topological superconductivity.

[*]*rchapai@anl.gov*

**INTRODUCTION**

Topological materials remain as a frontier in condensed matter physics, owing to their fascinating physical properties. Specifically, topological insulators and topological superconductors are being extensively explored due to their potential applications in quantum computation and spintronic technologies [1-6]. Topological superconductors are envisioned to exist in materials that possess topologically nontrivial bands and a superconducting ground state [6-12]. To achieve such exotic states, various approaches have been employed including doping topological insulators [13-15], applying external pressure to topological systems [16, 17], and fabricating heterostructures consisting of topological insulators and conventional superconductors [18-21]. However, superconductivity induced by chemical doping or pressure faces difficulties in achieving high superconducting volume fraction [17,19], and it can be challenging to fabricate an atomically sharp interface between two different materials [18-21]. An alternative approach to realizing topological superconductivity is to identify nontrivial topological bands within bulk superconductors [22-25], though materials in this category remain scarce [9, 26]. Only a few candidates, such as the non-centrosymmetric CePt$_3$Si [26] and orthorhombic UTe$_2$ [27] heavy fermion superconductors, and the layered transition-metal chalcogenide superconductors, like centrosymmetric PdTe$_2$ [28, 29] and non-centrosymmetric PbTaSe$_2$ [30], have shown promise. Joining these candidates, Kagome superconductors $A$V$_3$Sb$_5$ ($A$ = Cs, Rb, K) have recently been identified as having topologically non-trivial band structures with a non-zero Z$_2$ invariant [31, 32].

In parallel, the topological properties and Fermi surfaces within the parkerite and shandite family of compounds with composition $T_3M_2X_2$ ($T$ = Ni, Co, Rh, Pd, Pt; $M$ = In, Sn, Pb, Bi; $X$ = S, Se, Te) [33-40], have attracted considerable attention. Many of these ternary compounds, such as Rh$_3$Bi$_2$Se$_2$ and Rh$_3$Bi$_2$S$_2$, exhibit superconductivity and charge density wave (CDW) order [33, 34]. Other members, such as Co$_3$Sn$_2$S$_2$ and Pd$_3$Bi$_2$S$_2$, are known to host nontrivial topology [41-43]. However, the Fermi surface topology of superconducting parkerites remains poorly understood [33-40]. One such parkerite-type compound, Pd$_3$Bi$_2$Se$_2$, is a superconductor with $T_c \approx 1$ K with its superconducting properties reported in polycrystalline samples [35], thin films [37], and nanoparticles [38]. Recently, a non-zero Z$_2$ topological invariant of monoclinic parkerite-type Pd$_3$Bi$_2$Se$_2$ has been proposed theoretically [42-46], and unusual magneto-transport attributed to two-dimensional fermions has been reported in thin film samples [39]. A possible CDW transition has also been identified in low-pressure (~ 0.15 GPa) measurements [47]. Nevertheless, a detailed understanding of the Fermi surface, which is a prerequisite for comprehending correlated electronic behavior, has not yet been reported for $Pd_3Bi_2Se_2$ [33-40, 48, 49].

Here, we present the synthesis of large high-quality single crystals of $Pd_3Bi_2Se_2$ and detail its superconducting characteristics, normal state magneto-transport, and de Haas-van Alphen (dHvA) oscillations, to shed light on the electronic structure and Fermi surface (FS) topology. Superconductivity is observed in electrical resistivity with the onset of the transition at $T_c \approx 0.83$ K in zero field. The temperature dependence of the upper critical field, $H_{c2}(T)$ exhibits a modest anisotropy of less than 2, with zero-temperature values for the in-plane and out-of-plane directions of 9.7 mT and 5.2 mT, respectively with corresponding large Ginzburg-Landau (GL) coherence lengths of 190 nm and 111 nm. The temperature variation of the penetration depth, based on tunnel diode oscillator (TDO) measurements, is consistent with a complete superconducting gap. In the normal state, we observed clear dHvA oscillations at fields above 10 T and at low temperatures (<60 K), from which four frequencies are identified: $F_\alpha = 150$ T, $F_\beta = 293$ T, $F_\gamma = 375$ T and $F_\eta = 1017$ T, with the $F_\alpha$ frequency dominating the spectrum. Lifshitz-Kosevich analysis yields a small cyclotron effective mass ($m^* = 0.11 m_0$) and a nontrivial Berry phase ($\phi_B \approx \pi$) for the dominant orbit. These results are corroborated by density functional theory (DFT) calculations of the electronic structure, which reveal three doubly degenerate bands crossing the Fermi level. One of these bands results in a small, approximately ellipsoidal electron pocket centered on the $L_2$ point of the Brillouin zone with a cyclotron effective mass of $0.15 m_0$ for a magnetic field applied perpendicular to the *ab*-plane. The observed angular dependence of the $F_\alpha$ oscillations is well described using a model based on an ellipsoidal FS centered on the $L_2$ point and aligned with the D-$D_2$ direction in the Brillouin zone.

Multiple dHvA oscillation frequencies indicate multiple bands crossing the Fermi level. Indeed, the presence of multiple Fermi surface pockets is confirmed by the Hall resistivity measurements, which show nonlinear variation with the applied magnetic field. Despite the multiband nature, we observe non-saturating magnetoresistance that follows Kohler's scaling rule remarkably well, implying transport may be dominated by a single band or single scattering mechanism. Furthermore, our DFT calculations reveal a large three-dimensional Fermi surface sheet, supporting multiple open orbits that may account for the observed magneto-transport behavior and the overall low electronic anisotropy of $Pd_3Bi_2Se_2$.

**METHODS**

Single crystals of $Pd_3Bi_2Se_2$ were grown using the self-flux method. The starting materials, stoichiometric Pd powder (99.95%), Bi powder (99.99%) and Se powder (99.99) were mixed with excess Bi-Se (50% more by molar ratio), pressed into a pellet, and placed into an alumina crucible, that was then sealed in a fused silica tube under a vacuum of ~ 10 mTorr. The mixture was then heated to 610 °C and held for 24 hours, slowly cooled to 500 °C at a rate of 1 °C/h, and subsequently quenched to room temperature. Plate-like single crystals with typical size ~ $1.5 \times 1 \times 0.5$ mm$^3$ were obtained (see inset of Fig. S1(a) [50]). The structure of the as-grown crystals was determined through single crystal X-ray diffraction (XRD) measurements (BRUKER SMART APEX II, MoK$_\alpha$ radiation). Details of the structural refinement are provided in the Supplementary Materials [50]. The orientation of larger single crystals was determined through XRD measurement using a PANalytical X'Pert Pro diffractometer with CuK$_\alpha$ radiation. X-ray diffraction shows that our samples of $Pd_3Bi_2Se_2$ crystallize in the monoclinic space group $C2/m$ (#12) with refined lattice parameters $a = 11.7403(8)$ Å, $b = 8.4339(6)$ Å, $c = 8.4163(6)$ Å and $\beta = 133.834(1)°$, in good agreement with previous studies [47-49]. Electrical resistivity was measured using the standard four-probe technique in a Physical Properties Measurement System (PPMS-14T, Quantum Design) and in a dilution refrigerator (Bluefors) equipped with a 9-1-1 T triple-axis vector magnet. The TDO measurements were performed in a custom-built system operating at ~14.5 MHz in a $^3$He cryostat. In this technique, the change in the resonator frequency $\Delta f(T)$ is proportional to the change in magnetic susceptibility as $\Delta f(T) = -4\pi G\Delta\chi(T)$, where the geometrical factor G depends on the physical dimensions and geometry of the sample and resonator coil. In the Meissner state, the change in magnetic susceptibility reflects the change of the London penetration depth $\lambda(T)$ such that $\Delta f(T) \propto \Delta\lambda(T)$. The very small magnitude of the rf magnetic field (~20 mG) ensures that the sample remains fully in the Meissner state during measurements. Magnetic torque measurements were performed using piezo-resistive torque magnetometry with a 35 T resistive magnet at the National High Magnetic Field Laboratory (NHMFL) in Tallahassee, Florida. The samples were mounted on self-sensing cantilevers that were placed in a $^3$He cryostat. Measurements were performed with a balanced Wheatstone bridge that uses two piezoresistive paths on the cantilever as well as two resistors at room temperature that can be adjusted to balance the circuit. The voltage across the Wheatstone bridge was measured using a lock-in amplifier (Stanford Research Systems, SR860).

The electronic structure of $Pd_3Bi_2Se_2$ was computed using Density Functional Theory calculations. We employed the projector augmented wave method (PAW) as implemented in the Vienna Ab-initio Simulation Package (VASP) within the generalized gradient approximation (GGA) using the revised Perdew-Burke-Ernzerhof exchange-correlation functional for solids (PBEsol) [51, 52]. Potentials for each atom were chosen corresponding to the following electron configurations: Bi ($5d^{10}6s^26p^3$), Pd ($4d^95s^1$), and Se ($4s^24p^4$). First, optimization of the atomic positions, unit cell shape, and unit cell volume was performed on the primitive unit cell to a convergence level of $10^{-8}$ eV using an energy cutoff of 501 eV and including contributions from spin-orbit coupling and a Hubbard $U$ correction for the Pd $d$ states ($U_{Pd}$ = 2.19 eV). Geometry and unit cell parameters of the relaxed cell are given in Table S3 [50]. After relaxation, the charge density was first determined via a self-consistent calculation using a regular Γ-centered $k$-points mesh with the maximum distance between $k$-points set to 0.1 Å$^{-1}$. The density of states was determined using the same $k$-point mesh, the Fermi surface was calculated using a Γ-centered mesh of 1000 $k$-points spanning the Brillouin zone, and the band structure was determined along high symmetry $k$-point paths through the Brillouin zone. The c2x and XCrysDen software packages [53, 54] were utilized to visualize the Fermi surface, and the band structure was visualized using the pymatgen python library. Calculations of cyclotron effective masses and dHvA frequencies were derived from the VASP output using in-house code.

## RESULTS AND DISCUSSION

The main panel of Figure 1(a) shows the temperature dependence of the in-plane electrical resistivity, $ρ_{ab}(T)$, of $Pd_3Bi_2Se_2$, measured between 300 K and 1.8 K. $ρ_{ab}(T)$ exhibits typical metallic behavior, decreasing monotonically with temperature; $ρ_{ab}$ (1.8 K) ≈ 0.82 μΩ cm and the residual resistivity ratio (*RRR*), $ρ_{ab}$ (300 K)/$ρ_{ab}$ (1.8 K) ≈ 60. This *RRR* value is much larger than that reported earlier for polycrystalline [35] and thin film samples [37], and combined with low residual resistivity, reflects the high-quality of our single crystals. The upper inset of Fig. 1(a) displays $ρ_{ab}(T)$ between 1.8 K and 0.03 K measured in a dilution refrigerator. Superconductivity is observed in the electrical resistivity with a $T_{c,onset}$ of ≈ 0.83 K, and zero resistance is reached at ≈ 0.78 K. Note that the $T_c$ ≈ 0.80 K (defined as the mid-point of the transition) value of our single crystal sample is slightly lower than that reported for a polycrystalline sample (≈ 0.95 K) [35]. The normal state resistivity data can be fitted to $ρ_{ab}(T) = ρ_0 + AT^2$ below 40 K with $ρ_0$ = 0.76 μΩ cm and $A$ = 2.2 nΩ cm K$^{-2}$ (see lower inset of Fig. 1(a)). The value of $A$ falls within the range typically seen for correlated transition metals and A15 type superconductors [55, 56]. The $T^2$ dependence of $\rho_{ab}$ signals conventional Fermi-liquid ground state with dominant electron-electron scattering at low temperatures. $\rho_{ab}(T)$ above 40 K is approximately linear, reflecting dominant electron-phonon scattering at high temperatures.

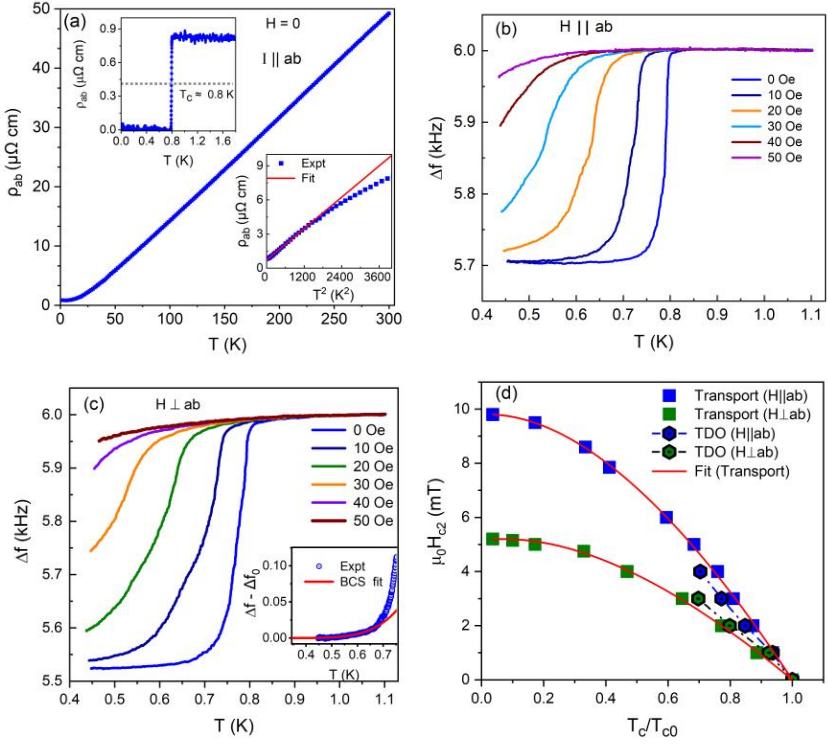

FIG. 1. (a) Temperature dependence of the electrical resistivity, $\rho_{ab}(T)$, of $Pd_3Bi_2Se_2$ measured between 300 K and 1.8 K. The upper inset displays $\rho_{ab}(T)$ measured between 1.8 K and 0.03 K. The horizontal dotted line represents the mid-point of the transition, used as a criterion for $T_c$. The lower inset shows $\rho_{ab}$ versus $T^2$. The solid line is a fit of the form $\rho = \rho(0) + AT^2$. The temperature dependence of the TDO frequency shift between 0.4 K and 1.1 K is shown as measured on a single crystal with $H \parallel ab$ (b), and $H \perp ab$ (c). Inset: Zero field data with BCS like fit. (d) $H_{c2}$ vs $T$ superconducting phase boundaries in reduced temperature $T_c/T_{c0}$ for both orientations and techniques. The solid lines are WHH model fits [57, 58].

The temperature dependence of the TDO frequency shift between 0.4 K and 1.1 K measured on a single crystal under $H \parallel ab$ is shown in Fig. 1(b). The sample displays superconductivity with the onset of the transition $T_c \approx 0.83$ K in zero field, which is consistent with the value observed in the electrical resistivity discussed above. With increasing field, the transition onset shifts to lower temperatures and is largely suppressed by 5 mT. As shown in Fig. 1(c), similar behavior is observed under $H \perp ab$. We attempt to analyze the zero-field temperature dependence TDO signal, $\Delta f(T)$. The solid line in the inset of Fig. 1(c) represents a fit of the data to the BCS weak-coupling behavior.

Although the accessible temperature range below $T_c$ is too small for a complete analysis of the low-temperature variation of the TDO signal, the available data suggest that $\Delta f(T)$ saturates at low temperatures, indicative of an exponential temperature dependence of $\lambda(T)$ and a complete superconducting gap.

To construct the complete superconducting phase diagram, we performed resistivity measurements in a dilution refrigerator following two protocols: with temperature sweep at fixed field and with field sweep at fixed low temperatures below $T_c$ (shown in Fig. S3[50]). From the data in Fig. 1(b-c) and Fig. S3, the superconducting phase diagram is constructed as shown in Fig. 1(d) (see Ref. [50] for the detailed criteria used to estimate $T_c$ and $H_{c2}$). A WHH fit [56-58] to the data yields the upper critical field $\mu_0 H_{c2}(0) \sim 9.7$ mT for in plane direction ($H \parallel ab$) and $\mu_0 H_{c2}(0) \sim 5.3$ mT for out of plane direction ($H \perp ab$), and upper critical field slopes at $T_c$ of -19.3 mT/K and -11.3 mT/K, respectively. The corresponding GL coherence lengths can be evaluated using the GL relations yielding large coherence lengths of $\xi_{out\ of\ plane}(0) \approx 111\ nm$ and $\xi_{in\ plane}(0) \approx 190\ nm$. We notice a modest anisotropy of the upper critical field, $\frac{H_{c2}^{ab}}{H_{c2}^{c}} \lesssim 2$, consistent with the three-dimensional Fermi surface observed through dHvA oscillation measurement and DFT calculation that we will discuss later. Our measured $\xi$ exceeds that reported for polycrystalline samples ($\approx 32\ nm$) [35] by more than a factor of 5 and is comparable to that of epitaxial thin films ($\approx 140\ nm$)[37]. Such evolution is expected since the superconducting coherence length increases with increasing purity (*RRR*), that is, increasing electron mean free path (*l*), as expressed by the dirty-limit relation $\xi \sim (l\xi_0)^{1/2}$ where $\xi_0$ is the BCS coherence length.

While there is no sign of any magnetic anomaly in $\rho_{ab}(T)$ (Fig. 1(a)), the electrical transport exhibits a large response to the application of a magnetic field. Figure 2(a) displays the transverse ($H \perp ab, I \parallel ab$) magnetoresistance $MR = \frac{[\rho(H) - \rho(0)]}{\rho(0)}$ measured at various temperatures. The $MR$ is positive at all temperatures and increases with decreasing temperature, reaching 300% at 1.8 K and 14 T, a value much larger than that reported for a thin film sample [39]. For a system with either a single band (or a dominant single band) or multiple bands with electron-hole compensation, the transverse $MR$ is expected to follow Kohler's scaling law [59, 60], provided that electron transport can be described by a single scattering time whose anisotropy does not change with temperature or field. In that case, all field and temperature dependent data collapse onto a single curve when plotted as $MR$ versus $\frac{H}{\rho(0)}$ described by the equation $MR = \mathcal{F}(H/\rho(0,T))$ [59, 60]. Here, $\mathcal{F}$ is a scaling function that reflects the electronic band structure. For small values of the argument $\frac{H}{\rho(0,T)}$, the $MR$ is typically quadratic, $MR \sim H^2$, while the high-field behavior depends on details of the electronic structure. In compensated materials, the $MR$ keeps growing as $H^2$. In contrast, in uncompensated materials with all closed electron orbits, the $MR$ in high fields saturates while open orbits induce saturation or a $H^2$-variation, depending on the details of the orbits. Polycrystalline samples, averaging over all types of orbits, display a linear high-field $MR$ [61, 62]. Figure 2(b) shows the magneto-transport data of our single-crystal $Pd_3Bi_2Se_2$ following Kohler's scaling rule remarkably well. As we will see, the DFT calculated electronic band structures described below indicate multiple electron and hole bands at the Fermi level. Therefore, the scaling property implies that magneto-transport is dominated by a single band and the overall dependence at 1.8 K is well described by $MR = 6.38(H/\rho(0))^{1.4}$, represented by the solid line in Fig. 2(b). In fields up to 14 T, we do not observe any signs of saturation, although in a log-log plot (Fig. S4 [50]) the slope of $MR$ vs $H$ in high fields is seen to decrease from 1.7 at low $H/\rho(0)$ to 1.4.

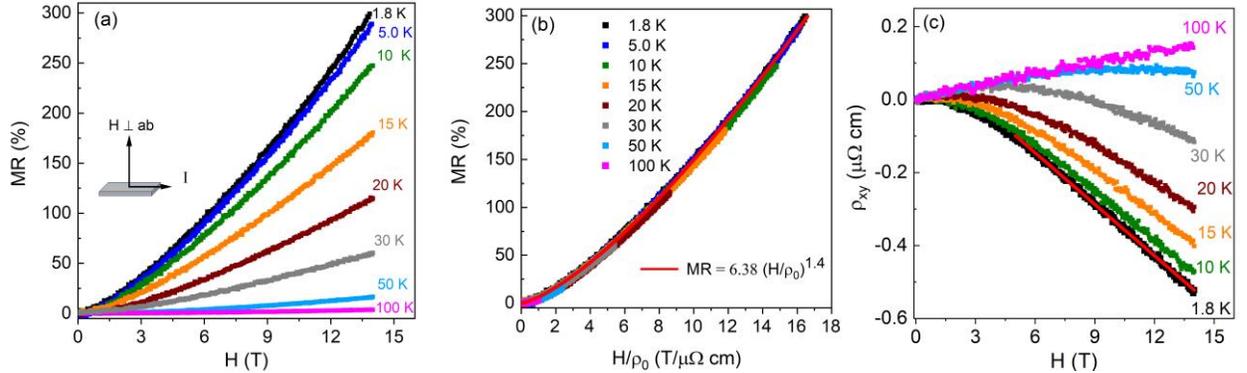

FIG. 2. (a) Transverse magnetoresistance of $Pd_3Bi_2Se_2$ at various temperatures. (b) Kohler scaling plot produced using the data in frame (a). Solid line (red) is a fit of the form $MR = 6.38(H/\rho(0))^{1.4}$ to the data at 1.8 K. (c) Hall resistivity ($\rho_{xy}$) measured at various temperatures. Solid line (red) is a linear fit to the data at 1.8 K in higher field regime (>5 T).

Figure 2(c) shows the field dependence of the Hall resistivity ($\rho_{xy}$) of $Pd_3Bi_2Se_2$ at different temperatures. While $\rho_{xy}$ varies linearly with $H$ with a positive slope at high temperature and low field, a nonlinearity in $\rho_{xy}(H)$ gradually develops and becomes obvious above 7 T at 50 K, leading to a sign change at high $H$. Generally, a nonlinear Hall resistivity with a sign change indicates that electron and hole bands with unequal carrier concentrations and mobilities both contribute to transport [61]. Accordingly, we analyzed the magneto-transport data using the standard two-band model [61, 62]. Even though this model contains extensive assumptions and simplifications, it is nevertheless useful for exploring trends in transport behavior. While the non-saturating magnetoresistance following Kohler's scaling rule could indicate electron-hole compensation ($n_e = n_h$), the non-linear sign-changing Hall resistivity is not consistent with compensation. Then, the limiting linear variation of $\rho_{xy}$ at 1.8 K and high fields (red solid line in Fig. 2(c)) signals the approach to the high-field limit where the Hall coefficient ($R_H$) is given through $(n_h - n_e) = 1/eR_H = -1.3 \times 10^{22}\ cm^{-3}$. Thus, this model predicts $n_e > n_h$ in the system. Together with the additional constraints given by the experimental values of the zero-field conductivity, $\sigma_0 = e(n_e\mu_e + n_h\mu_h)$, and the Hall resistivity, $\rho_{xy}/H = e(n_h\mu_h^2 - n_e\mu_e^2)/\sigma_0^2$, we deduce an overall charge density of order $10^{22}\ cm^{-3}$. While this value is near the low end of electron concentrations typically seen in good metals, it is nevertheless consistent with the metallic behavior of Pd$_3$Bi$_2$Se$_2$ observed in the resistivity measurement (Fig. 1(a)). However, we note that within the two-band ansatz, the approach to the high-field limit in $\rho_{xy}$ and the simultaneous absence of saturation in $\rho_{xx}$ in the same field range are inconsistent. We therefore conclude that the magneto-transport data in Fig. 2 cannot be accounted for within the two-band model in its standard form. One of the assumptions of this model relates to the absence of open orbits [60, 61]. The inclusion of such orbits will, depending on their geometry, induce non-saturating magnetoresistance and an additional contribution to the Hall resistivity, such that the standard relation with the carrier concentration is no longer valid [61-63], in agreement with our observed magneto-transport data.

To explore the possibility of open orbits, we performed DFT calculations of the electronic band structure of Pd$_3$Bi$_2$Se$_2$. The band structure presented in Fig. 3(a) reveals three pairs of doubly degenerate bands crossing the Fermi level. As shown in Fig. 3(b) and 3(c), bands #143 forms two large pockets centered at $\Gamma$ and M$_2$, and band #145 consists of a complex continuous three-dimensional sheet. In particular, the Fermi surface sheet associated with band #145 supports the notion of open orbits in this system, with infinite channels running along both the *a* and *c\** directions (Fig. S9 [50]). These open orbits may thus contribute to the observed magnetoresistance described above. Band #147 crosses the Fermi level ($E_F$) twice: once at the L$_2$ point of the Brillouin zone and again at a non-high symmetry point along the path between $\Gamma$ and Y$_4$. These crossings manifest as two closed electron Fermi surface pockets in the Brillouin zone shown in Fig. 3(d). The bands associated with these electron pockets are highly dispersed and nearly parabolic close to $E_F$, so the resulting Fermi surface is very sensitive to the precise value of the Fermi energy [48]. For instance, with the magnetic field $H$ oriented along $c^*$ (perpendicular to the crystal layers), DFT yields a dHvA frequency of 14 T for the Fermi surface pocket at $L_2$. However, raising $E_F$ by ~0.10 eV (dashed line in Fig. 3(a)) enlarges the $L_2$ pocket and brings the calculated frequency in line with the experimental value ($F_\alpha \approx 145$ T).

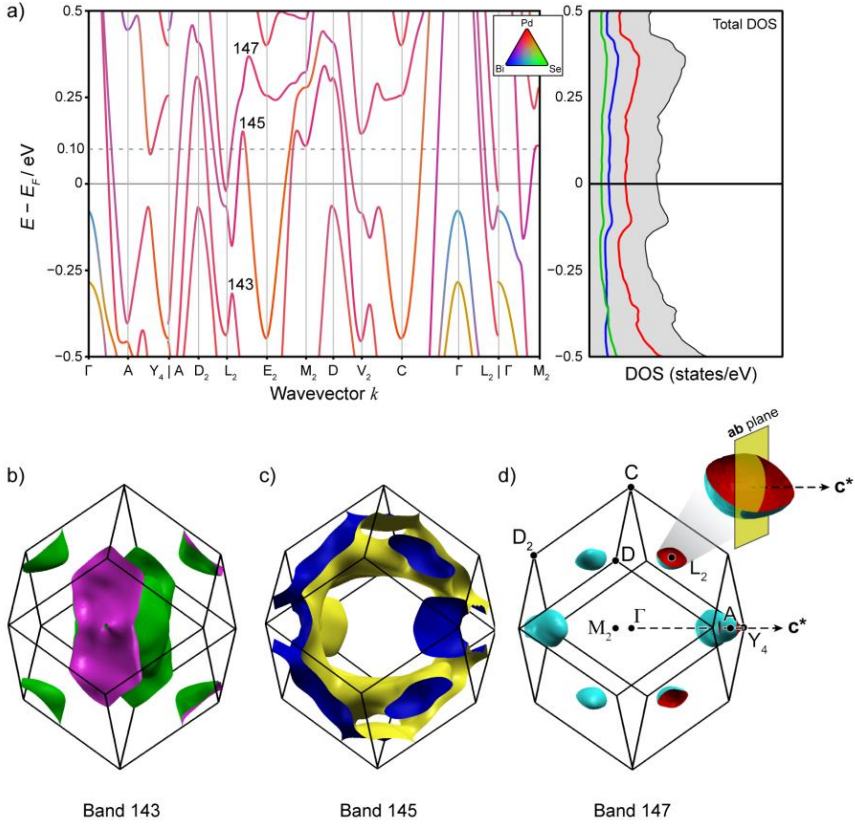

FIG. 3. (a) Electronic band structure and density of states (DOS) for $Pd_3Bi_2Se_2$ in the presence of spin-orbit coupling. The DOS from Pd (red), Bi (blue), Se (green) and the sum. Raising the Fermi level by 0.10 eV slightly expands the maximal orbit of the electron pocket at $L_2$, bringing it into alignment with the experimentally measured $F_\alpha$. (b-d) Fermi surfaces corresponding to three doubly degenerate bands that cross the Fermi level presented in frame (a).

In concert, a cyclotron effective mass of $0.15m_0$ ($m_0$ is the free electron mass) is calculated, in good agreement with the measured $m^* = (0.11 \pm 0.02)m_0$, discussed later. The Fermi surfaces shown in Figs. 3(b), 3(c), and 3(d) are calculated using the shifted Fermi level, i.e., electron doping. A plausible source for such electron doping could be $Se^{2-}$ vacancies introduced during synthesis [37, 64]. Assuming a rigid band model of the electronic structure, a shift of 0.10 eV amounts to 0.2 electrons per formula unit, corresponding to a ≈ 5% Se deficiency. Neither single crystal X-ray diffraction (Table S2), energy-dispersive X-ray spectroscopy (Fig. S2), nor inductively coupled plasma spectroscopy (Table S4, [50]) reveal Se-deficiency at this scale, although such non-stoichiometry may lie at the detection limit of these techniques. It should also be noted that the assumptions inherent to DFT may not precisely capture the nuances of the band structure near $E_F$, introducing a quantitative error in the estimated doping level. Further theoretical work and experimental validation are desirable for a comprehensive understanding of these discrepancies. Nevertheless, the angular dependence of our measured dHvA frequency (see below) and a good agreement with the cyclotron effective mass allow us to identify the L$_2$ pocket as the origin of $F_\alpha$, as we will discuss in detail below.

To obtain experimental signatures of the Fermi surface of Pd$_3$Bi$_2$Se$_2$, we measured quantum oscillations using highly sensitive torque magnetometry at low temperatures and high magnetic fields up to 35 T. Figure 4(a) displays the field dependence of the magnetic torque, $\tau(H)$, measured at 0.5 K in a field applied approximately perpendicular to $ab$ plane (parallel to $c^*$) (also see Fig. S7, [50]). dHvA oscillations are visible above $\approx$ 10 T. By subtracting a smooth polynomial background, we extract the oscillatory part of the magnetic torque, $\Delta\tau$, as shown in the inset of Fig. 4(a). Figure 4(b) shows $\Delta\tau$ plotted as a function of the inverse magnetic field ($H^{-1}$) at various temperatures, revealing the rapid suppression of the oscillation amplitude with increasing temperature. Nevertheless, the oscillations persist up above 60 K, allowing us to determine the carrier cyclotron effective mass as discussed later. From the Fast Fourier Transform (FFT) performed in the field range of 15-35 T, four oscillation frequencies are identified as $F_\alpha = (150 \pm 26)$ T, $F_\beta = (293 \pm 10)$ T, $F_\gamma = (375 \pm 20)$ T and $F_\eta = (1017 \pm 12)$ T, shown in Fig. 4(c) and Fig. S5 [50] with the low frequency dominating the spectrum. The error bars are determined by the half-width at half-height of the FFT peaks. The dominant frequency, $F_\alpha = 150$ T, occupies a small portion ($\approx$ 1.3%) of the total area of the primitive in-plane Brillouin zone of monoclinic Pd$_3$Bi$_2$Se$_2$.

The dHvA oscillations can be described by the Lifshitz-Kosevich formula [65, 66]

$$\Delta\tau \propto -\frac{1}{F}\frac{\partial F}{\partial \theta} H^\lambda R_T R_D R_S \sin\left[2\pi\left\{\frac{F}{H} - \left(\frac{1}{2} - \Phi\right)\right\}\right] \quad (1)$$

where $F$ (in units of Tesla) is the frequency of the oscillation, $\theta$ is the field angle, $R_T = \dfrac{A\left(\frac{m^*}{m_0}\right)\frac{T}{H}}{sinh\left(A\left(\frac{m^*}{m_0}\right)\frac{T}{H}\right)}$ is the thermal damping factor, $A = \dfrac{2\pi^2 k_B m_0}{e\hbar} = 14.7$ T/K, $R_D = exp\left(-A\dfrac{m^*}{m_0}\dfrac{T_D}{H}\right)$ is the Dingle damping factor ($T_D$ is the Dingle temperature), and $R_S = \cos\left(\pi g^* \frac{m^*}{2m_0}\right)$ is the spin reduction factor ($m^*$ is the cyclotron effective mass and $g^*$ is the effective g-factor). The exponent $\lambda = 1$ in Eq. (1) is chosen for a two-dimensional (2D) FS, and $\lambda = 3/2$ for a three-dimensional (3D) FS [67]. In addition, the phase factor, $\Phi = \frac{\Phi_B}{2\pi} + \delta$, where $\Phi_B$ is the Berry phase, and $\delta = 0$ for a 2D and $\pm 1/8$ for a 3D FS (+/- sign corresponds to the minima (+)/maxima (-) of the cross-sectional area of the FS for the case of an electron band; for a 3D hole band, the sign of $\delta$ is opposite [68]. The quantum oscillation frequency is related to the area ($S$) of the extremal orbit through the Onsager relation, $F = \left(\frac{\hbar}{2\pi e}\right)S$.

The temperature dependence of the FFT amplitude of the α band is displayed in the inset of Fig. 4(c). From a fit of the thermal damping factor $R_T$ (defined in Eq. (1)) to these data, we obtain a light cyclotron effective mass of $m_\alpha^* = (0.11 \pm 0.02)m_0$. This is in good agreement with $m_\alpha^* = 0.15m_0$ derived from our DFT calculation, corroborating our assignment of the $L_2$-pocket as discussed earlier. Owing to their approximate 1:2 ratio, the $F_\beta$ frequency ($F_\beta = 293$ T) may be the second harmonic of $F_\alpha$. The amplitude of this oscillation frequency is too small to confirm this hypothesis by extracting an effective mass, which would be expected to be twice that of the fundamental mass $m_\alpha^*$ (i.e. $m_\beta^* = 0.22m_0$). Our DFT calculations reveal that the area of the electron pocket near $Y_4$ corresponds to a dHvA frequency of 289 T, providing an alternative explanation for the $F_\beta$ frequency. Notably, the closed Fermi surface branches originating from band #143 produce dHvA frequencies greatly exceeding any of our measured values.

The field dependence of the oscillation amplitude at a given temperature can provide information about the Dingle temperature, $T_D$, through the analysis of Dingle plots (see Fig. S6, [50]). From $T_D$ ($\approx 37$ K), the quantum relaxation time, $\tau_q$, can be estimated through the relation $\tau_q = \frac{\hbar}{2\pi T_D k_B} = 0.32 \times 10^{-13}$ sec, which, in turn, is proportional to the quantum mobility $\mu_q = \frac{e\tau_q}{m^*} = 0.05$ m²/V sec. Additional parameters obtained from the analysis of the dHvA oscillations are listed in Table S5 [50]. Within the Drude model, the conductivity associated with the $L_2$-pockets can be estimated using the parameters obtained from the dHvA measurements as $\sigma = en\mu$, where $\mu$ is the mobility given above and $n$ is the volume concentration [69] of electrons per $L_2$-pocket given as $n = \frac{2\sqrt{2}}{3\pi^2}\left(\frac{e}{\hbar}\right)^{\frac{3}{2}} F_{min}^{1/2} F_{max} \approx 1.5 \times 10^{19}$ cm⁻³, resulting in a resistivity of ~395 μΩ cm for the two L$_2$ pockets in the Brillouin zone. Here, $F_{min}$ ~ 116 T and $F_{max}$ ~260 T are the minimum and maximum cyclotron frequency as determined within an ellipsoidal model from angular dependent dHvA measurements as described below. Even though their mobility is high, the contribution of the L$_2$-pockets to the electric transport is small, chiefly because of their small size. These estimates suggest that magneto-transport is dominated by the large Fermi surface sheets shown in Figs. 3(b) and 3(c).

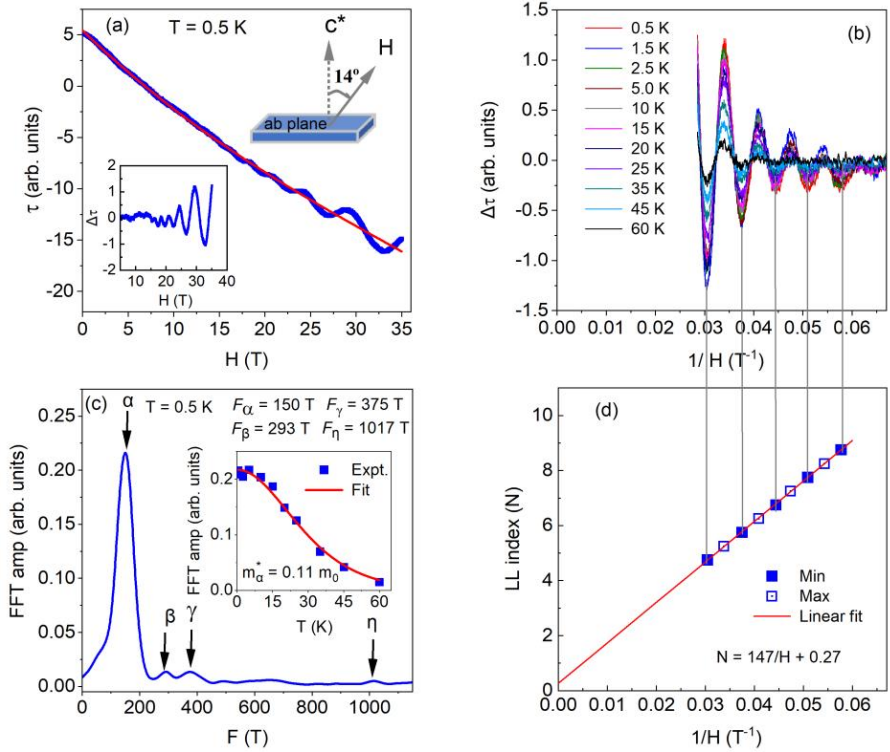

FIG. 4. (a) Field dependence of the magnetic torque $\tau$ ($H$) of Pd$_3$Bi$_2$Se$_2$ at T = 0.5 K. The field is applied at an angle $\theta$ ~ 14° from the $c^*$ axis as depicted in the upper inset. This tilt angle was chosen for the temperature dependent measurement as the oscillations are more pronounced at $\theta$ ~ 14°, and is not far from H $\perp$ $ab$-plane [50]. The red curve is a third order polynomial fit to the data. Lower inset: background subtracted $\Delta\tau$ versus $H$. (b) $\Delta\tau$ plotted as a function of $H^{-1}$ at various temperatures. (c) FFT amplitude as a function of frequency at 0.5 K. Inset: Variation of FFT amplitude with temperature for the α band. The solid line is a fit with thermal damping term of Eq. (1). (d) Landau level fan plot constructed for the dominating α band. The red line is a linear fit to the data.

Phase analysis of the dHvA oscillations can reveal the topological properties of the associated carriers. For such analysis, the Berry phase can be extracted from the Landau level (LL) fan diagram [5, 65, 67, 70]. Since the dHvA spectrum is dominated by a single low frequency (α band), the LL fan diagram can be constructed unambiguously without filtering the signal [71, 72].

Since the magnetic torque is proportional to the density of states at the Fermi level [65, 67], the LL fan diagram is constructed by assigning the oscillation minima in $\Delta\tau$ to $N$-1/4 and maxima to $N$ +1/4, where $N$ is the Landau level index [65]. As shown in Fig. 4(d), $N(H^{-1})$ can be fitted with $N = 0.27 + 147/H$. The y-axis intercept of the linear fit, following Eq. (1), yields the Berry phase as $\frac{\Phi_B^\alpha}{2\pi} \sim 0.4$, close to the value of 0.5 expected for a topological orbit. In this estimate, we used $\delta = -\frac{1}{8}$, appropriate for a 3D maximal electron orbit as discussed above. The slope of the linear fit corresponds to the frequency $F_\alpha = 147$ T, in good agreement with that obtained from the FFT spectra.

The Topological Materials Database [42, 44-46] identifies $Pd_3Bi_2Se_2$ as a $Z_2$ topological material, which is characterized by a continuous direct gap at each $k$-point in the Brillouin zone. The observed π-phase would thus be consistent with the non-trivial topological nature of the α band. Furthermore, the Landau level fan diagrams constructed for various angles reveal an angular dependence of the phase factor for the α band (Fig. S8 (a-b) [50]), which may signal a topological phase transition [73-76]. Such interpretations should be taken with caution, however, as it has been noted that additional contributions to the phase shift of the quantum oscillations can arise from orbital momentum and Zeeman coupling [77]. Since these other contributions are presently not known for $Pd_3Bi_2Se_2$, a definite statement on the change of the topological state of the α-band requires further studies.

The angular dependence of quantum oscillations can provide further information about the Fermi surface topology. Figure 5(a) displays $\Delta\tau$ ($H$) versus $H^{-1}$ of $Pd_3Bi_2Se_2$ measured at $T = 0.5$ K for different angles $\theta$ (defined in the inset in frame (c)). Upon the variation of $\theta$, the dHvA oscillations change in both amplitude and peak position. Through FFT analysis, shown in Fig. 5(b), we can systematically track the dominant α band for various measured angles. Unfortunately, the other bands could not be unambiguously tracked due to their much smaller amplitudes.

The FFT amplitude of the α band displays a distinct angular dependence such that for $\theta > 60°$ it can no longer be identified unambiguously. Several factors may contribute to such angular variation. The torque-technique introduces angular dependencies of the oscillation amplitude stemming from the fact that this technique probes the magnetization component that is transverse to the applied field direction. These are contained in the factor $\frac{1}{F}\frac{dF}{d\theta}$ in Eq.(1) [65], which for an ellipsoid is given as $-\frac{1}{F}\frac{dF}{d\theta} = -\frac{(1-\Gamma^2)\cos(\theta)\sin(\theta)}{\cos^2(\theta)+\Gamma^2\sin^2(\theta)}$ with $\Gamma = \frac{F_{min}}{F_{max}}$ the anisotropy of the dHvA frequency. This relation yields the expected vanishing torque at the symmetry points at $\theta = 0°$ and $\theta = 90°$, and a rapid decrease of the torque at angles larger than $\approx 60°$. This drop becomes more precipitous, the larger the anisotropy is, that is, the smaller $\Gamma$ is. Furthermore, frequently a strong suppression of dHvA oscillations on layered materials is found for the in-plane field orientation [78, 79]. A possible reason is related to the reduced electron mobility in the inter-layer direction which would induce an increased Dingle temperature at large angle and the corresponding exponential suppression of the oscillations.

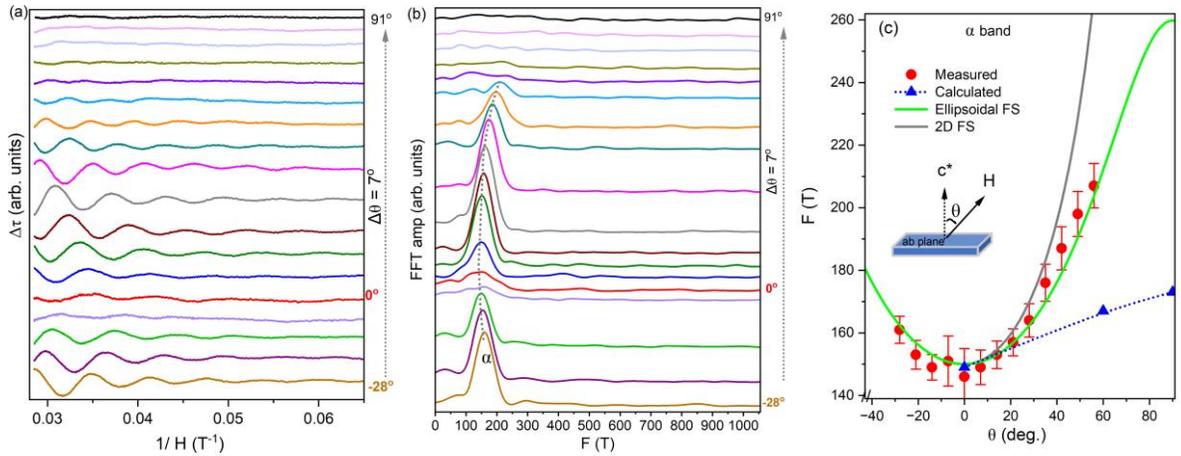

FIG. 5. (a) $\Delta\tau(H)$ vs. $H^{-1}$ of $Pd_3Bi_2Se_2$ measured at $T = 0.5$ K and at different angles $\theta$ (defined in the inset of frame (c)). A vertical shift is added to the data for visual clarity. These data are obtained after subtracting a smooth background from the field dependent torque data, $\tau(H)$ (Fig. S7 [50]). In the background subtracted data, the change in frequency and the phase of the oscillations are more apparent. (b) The FFT spectrum obtained from the data in frame (a). The dashed line is a visual guide, indicating the progression of the α band. (c) Angular dependence of $F_\alpha$ measured (red). The error bars are taken as the half-width at half-height of the FFT peaks. Solid lines represent 2D FS (gray), 3D ellipsoidal FS (green) and the DFT calculated values (blue) as described in the text. Inset: A sketch defining the field angle direction.

Figure 5(c) summarizes the angular dependence of $F_\alpha$. The measured dHvA frequency increases symmetrically for positive and negative angles implying that the FS pocket is symmetric about the plane spanned by the zero-angle field direction and the rotation axis. Referring to the geometry of the Brillouin zone in Fig. 3(d) and Fig. S10 [50], this finding is consistent with the *a*-axis of the conventional base-centered monoclinic cell being the rotation axis. In *k*-space, this axis corresponds to the $\Gamma$-$M_2$ direction. By the same reasoning, the $Y_4$ pocket has an irregular shape (see Fig. 3(d)) such that no rotation will yield a symmetric angular dependence of the oscillation frequency, excluding this pocket as the source of $F_\alpha$. The observed increase of $F_\alpha$ with angle falls short of the $\frac{1}{\cos(\theta)}$ variation of a 2D FS but is consistent with a FS pocket in the shape of an ellipsoid. In fact, a fit for an ellipsoidal FS pocket aligned with the D-D$_2$ direction and rotated around the Γ-M$_2$ axis (Fig. 5(c) and Fig. S10 [50]) describes the data well with an anisotropy of Γ ≈ 0.45. We note that the rotation axis is not aligned with a principal axis of the ellipsoid but subtends an angle of approximately 45°. Therefore, the minimum frequency is $F_{\min}$ ≈ 116 T, and not the frequency measured at $\theta = 0°$, which is 145 T. Also included in Fig. 5(c) are the dHvA frequencies calculated using DFT with $E_F$ shifted by ≈ 0.10 eV. While this calculation underestimates the angular dependence of $F_\alpha$ by approximately 20% at 60°, it qualitatively reproduces the monotonic increase in frequency with increasing $\theta$.

## SUMMARY

In summary, we report on the growth of single-crystal samples of Pd$_3$Bi$_2$Se$_2$ and present their superconducting and normal state magneto-transport properties. Superconductivity is observed in electrical resistivity and TDO measurements, with a transition temperature at $T_c$ ≈ 0.80 K. The availability of large bulk single crystals enables the study of their anisotropic electronic properties. The samples are of high purity, as evidenced by a large residual resistivity ratio of ≈ 60 and the observation of dHvA oscillations in fields as low as 10 T. Our magnetoresistance data exhibit Kohler-type scaling behavior, with an unusual $H^{1.4}$ dependence, suggesting contributions from multiple bands, as reflected in the Hall resistivity with nonlinear field dependence. Through analysis of the observed dHvA oscillations, we identify four bands with frequencies of $F_\alpha = 150$ T, $F_\beta = 293$ T, $F_\gamma = 375$ T and $F_\eta = 1017$ T. Among these bands, the α band (150 T) is the most dominant.

Angular dependent dHvA measurements, in conjunction with DFT calculations allow us to assign the α-band to a nearly ellipsoidal electron pocket centered on the L$_2$ point of the Brillouin zone. The calculated dHvA frequency of the α-pocket originating form band #147 can be aligned with our experimental results with an upward shift of the Fermi level by 0.10 eV, implying light electron doping of the sample by 0.2 electrons per formula unit. This ellipsoidal pocket displays modest anisotropy in the angular dependence of the dHvA oscillations. Lifshitz-Kosevich analysis of the dHvA oscillations reveals a small cyclotron effective mass $m^* = (0.11 \pm 0.02)m_0$ for this electron pocket, in good agreement with the value of $0.15m_0$ obtained from DFT calculations. The DFT calculations further reveal a large, three-dimensional extended Fermi surface sheet originating from band #145. This Fermi surface supports open orbits in the *ab*-plane of the crystal and may account for the observed magneto-transport behavior and the overall low electronic anisotropy of $Pd_3Bi_2Se_2$. For instance, the anisotropy of the upper critical field measured in fields parallel and perpendicular to the *ab*-planes is found to be less than two. Landau fan diagrams constructed from the dHvA oscillations yield a nontrivial Berry phase for the dominant orbit, indicating nontrivial band topology. The nontrivial topology present in the bulk superconductor $Pd_3Bi_2Se_2$ makes it a promising candidate for exploring potential topological superconductivity.

## ACKNOWLEDGMENTS

Acknowledgement: This work was supported by the U.S. Department of Energy, Office of Science, Basic Energy Sciences, Materials Sciences and Engineering. DFT calculations were performed using resources provided by the Center for Nanoscale Materials, a U.S. Department of Energy Office of Science User Facility under Contract No. DE-AC02-06CH11357. DFT calculations presented here are based upon work supported by Laboratory Directed Research and Development (LDRD) funding from Argonne National Laboratory, provided by the Director, Office of Science, of the U.S. Department of Energy under Contract No. DE-AC02-06CH11357. The EDX measurement performed at the Center for Nanoscale Materials, a U. S. Department of Energy, Office of Science User Facility, was supported by the U. S. DOE, Office of Basic Energy Sciences, under Contract No. DE-AC02-06CH11357. We thank Hengdi Zhao for help with the EDX measurements. A portion of this work was performed at the National High Magnetic Field Laboratory, Tallahassee which is supported by National Science Foundation Cooperative Agreement No. DMR-2128556 and the State of Florida.

## REFERENCES

[1] L. Fu, C. L. Kane, and E. J. Mele, Topological insulators in three dimensions, *Phys. Rev. Lett.* **98**, 106803 (2007).

[2] Y. L. Chen, J. G. Analytis, J.-H. Chu, Z. K. Liu, S.-K. Mo, X. L. Qi, H. J. Zhang, D. H. Lu, X. Dai, Z. Fang *et al.*, Experimental realization of a three-dimentional topological insulator, $Bi_2Te_3$, *Science* **325**, 178 (2009).

[3] Y. Xia, D. Qian, D. Hsieh, L. Wray, A. Pal, H. Lin, A. Bansil, D. Grauer, Y. S. Hor, R. J. Cava and M. Z. Hasan, Observation of a large-gap topological-insulator class with a single Dirac cone on the surface, *Nat. Phys.* **5**, 398 (2009).

[4] X.-L. Qi and S.-C. Zhang, Topological insulators and superconductors, *Rev. Mod. Phys.* **83**, 1057 (2011).

[5] Y. Ando, Topological insulator materials, *J. Phys. Soc. Jpn*, **82**, 102001 (2013).

[6] L. Fu and C. L. Kane, Topological insulators with inversion symmetry, *Phys. Rev.* B **76**, 045302 (2007).

[7] C. Beenakker and L. Kouwenhoven, A road to reality with topological superconductors, *Nat. Phys.* **12**, 618 (2016).

[8] R. M. Lutchyn, J. D. Sau, and S. D. Sarma, Majorana Fermions and a topological phase transition in semiconductor-superconductor heterostructures, *Phys. Rev. Lett*. **105**, 077001 (2010).

[9] M. Sato and Y. Ando, Topological superconductors: a review, *Rep. Prog. Phys*. **80**, 076501 (2017).

[10] J. C. Y. Teo and C. L. Kane, Majorana Fermions and non-abelian statistics in three dimensions, *Phys. Rev. Lett*. **104,** 046401 (2010).

[11] X. -L. Qi, T. L. Hughes, S. Raghu, and S.-C. Zhang, Time-reversal-invariant topological superconductors and superfluids in two and three dimensions, *Phys. Rev. Lett*. **102**, 187001 (2009).

[12] S. Deng, L. Viola, and G. Ortiz, Majorana modes in time-reversal invariant *s*-wave topological superconductors, *Phys. Rev. Lett*. **108**, 036803 (2012).

[13] Y. S. Hor, A. J. Williams, J. G. Checkelsky, P. Roushan, J. Seo, Q. Xu, H. W. Zandbergen, A. Yazdani, N. P. Ong and R. J. Cava, Superconductivity in $Cu_xBi_2Se_3$ and its implications for pairing in the undoped topological insulator, *Phys. Rev. Lett*. **104**, 057001 (2010).

[14] M. P. Smylie, K. Willa, H. Claus, A. Snezhko, I. Martin, W.-K. Kwok, Y. Qiu, Y. S. Hor, E. Bokari, P. Niraula, A. Kayani, V. Mishra and U. Welp, Robust odd-parity superconductivity in the doped topological insulator $Nb_xBi_2Se_3$, *Phys. Rev. B* **96**, 115145 (2017).

[15] L. A. Wray, S.-Y. Xu, Y. Xia, Y. S. Hor, D. Qian, A. V. Fedorov, H. Lin, A. Bansil, R. J. Cava and M. Z. Hasan, Observation of topological order in a superconducting doped topological insulator, *Nat. Phys*. **6**, 855 (2010).

[16] J. L. Zhang, S. J. Zhang, H. M. Weng, W. Zhang, L. X. Yang, Q. Q. Liu, S. M. Feng, X. C. Wang, R. C. Yu, L. Z. Cao *et al*., Pressure-induced superconductivity in topological parent compound $Bi_2Te_3$, *Proc. Natl. Acad. Sci. USA* **108**, 24 (2011).

[17] H. Takahashi, A. Sugimoto, Y. Nambu, T. Yamauchi, Y. Hirata, T. Kawakami, M. Avdeev, K. Matsubayashi, F. Du, C. Kawashima *et al*., Pressure-induced superconductivity in the iron-based ladder material $BaFe_2S_3$, *Nat. Mater*. **14**, 1008 (2015).

[18] B. Jäck, Y. Xie, J. Li, S. Jeon, B. A. Bernevig, A. Yazdani, Observation of a Majorana zero mode in a topologically protected edge channel, *Science* **364**, 1255 (2019).

[19] G. -H. Lee and H. -J. Lee, Proximity coupling in superconductor-graphene heterostructures, *Rep. Prog. Phys*. **81**, 056502 (2018).

[20] R. M. Lutchyn, J. D. Sau, and S. D. Sarma, Majorana fermions and a topological phase transition in semiconductor-superconductor heterostructures, *Phys. Rev. Lett*, **105**, 077001 (2010).

[21] H. Yi, Y. F. Zhao, Y. T. Chan, J. Cai, R. Mei et al, Interface-induced superconductivity in magnetic topological insulators, *Science* **383**, 634 (2024).

[22] M. S. Bahramy, O. J. Clark, B.-J. Yang, J. Feng, L. Bawden, J. M. Riley, I. Markovic, F. Mazzola, V. Sunko, D. Biswas *et al*., Ubiquitous formation of bulk Dirac cones and topological surface states from a single orbital manifold in transition-metal dichalcogenides, *Nat. Mater*. **17**, 21 (2018).

[23] C. Xu, N. Wu, G-X. Zhi, B.-H. Lei, X. Duan, F. Ning, C. Cao and Q. Chen, Coexistence of nontrivial topological properties and strong ferromagnetic fluctuations in quasi-one-dimensioal $A_2Cr_3As_3$, *npj Comput. Mater.* **30** 6 (2020).

[24] P. Zhang, Z. Wang, X. Wu, K. Yaji, Y. Ishida, Y. Kohama, G. Dai, Y. Sun, C. Bareille, K. Kuroda *et al*., Multiple topological states in iron-based superconductors, *Nat. Phys*. **15**, 41 (2019).

[25] L. Fu, and C. L. Kane, Superconducting proximity effect and Majorana fermions at the surface of a topological insulator, *Phys. Rev. Lett*. **100,** 096407 (2008).

[26] M. Mandal, N. C. Drucker, P. Siriviboon, T. Nguyen, A. Boonkird, T. N. Lamichhane, R. Okabe, A. Chotrattanapituk and M. Li, Topological superconductors from a materials perspective, *Chem. Mater. **35,*** 6184 (2023).

[27] L. Jiao, S. Howard, S. Ran, Z. Wang, J. Olivares Rodriguez *et al*., Chiral superconductivity in heavy-fermion metal $UTe_2$, *Nature* **579**, 523 (2020).

[28] H.-J. Noh, J. Jeong, E.-J. Cho, K. Kim, B. I. Min, B.-G. Park, Experimental realization of Type-II Dirac Fermions in a $PdTe_2$ superconductor, *Phys. Rev. Lett*. **119**, 016401(2017).

[29] F. Fei, X. Bo, R. Wang, B. Wu, J. Jiang, D. Fu, M. Gao, H. Zheng, Y. Chen, X. Wang *et al.,* Nontrivial Berry phase and type-II Dirac transport in the layered material $PdTe_2$, *Phys. Rev*. B **96**, 041201 (R) (2017).

[30] T.-R. Chang, P.-J. Chen, G. Bian, S.-M. Huang, H. Zheng *et al*., Topological Dirac surface states and superconducting pairing correlations in $PbTaSe_2$, *Phys. Rev*. B **93**, 245130 (2016).

[31] B. R. Ortiz, S. M. L. Teicher, Y. Hu, J. L. Zuo, P. M. Sarte, E. C. Schueller, A. M. M. Abeykoon, M. J. Krogstad, S. Rosenkranz, R. Osborn *et al*., $CsV_3Sb_5$: A $Z_2$ Topological kagome metal with a superconducting ground state, *Phys. Rev. Lett*. **125**, 247002 (2020).

[32] S. D. Wilson and B. R. Ortiz, $AV_3Sb_5$ kagome superconductors, *Nat. Rev. Mater*. (2024). https://doi.org/10.1038/s41578-024-00677-y

[33] T. Sakamoto, M. Wakeshima, Y. Hinatsu and K. Matsuhira, Charge-density-wave superconductor $Bi_2Rh_3Se_2$, *Phys. Rev. B* **75**, 060503 (R) (2007).

[34] U. S. Kaluarachchi, W. Xie. Q. Lin, V. Taufour, S. L. Budko, G. J. Miller, and P. C. Canfield, Superconductivity versus structural phase transition in the closely related $Bi_2Rh_{3.5}S_2$ and $Bi_2Rh_3S_2$, *Phys. Rev*. B **91**, 174513 (2015).

[35] T. Sakamoto, M. Wakeshima, Y. Hinatsu and K. Matsuhira, Transport properties in normal-metal $Bi_2Pd_3S_2$ and superconducting $Bi_2Pd_3Se_2$, *Phys. Rev. B* **78**, 024509 (2008).

[36] T. Sakamoto, M. Wakeshima, and Y. Hinatsu, Superconductivity in ternary chalcogenides $Bi_2Ni_3X_2$ (X= S, Se), *J. Phys.: Condens. Matter.* **18**, 4417 (2006).

[37] J. Lapano, Y. Y. Pai, A. R. Mazza, J. Zhang, T. Isaacs-Smith, P. Gemperline, L. Zhang, H. Li, H. N. Lee, G. Eres *et al*., Self-regulated growth of candidate topological superconducting parkerite by molecular beam epitaxy, *APL Mater*. **9**, 101110 (2021).

[38] M. Roslova, L. Opherden, I. Veremchuk, L. Spillecke, H. Kirmse, T. Herrmannsdörfer, J. Wosnitza, T. Doert. and M. Ruck, Downscaling effect on the superconductivity of $Pd_3Bi_2X_2$ (X= S or Se) nanoparticles prepared by microwave-assisted polyol synthesis, *Inorg. Chem.* **55**, 8808 (2016).

[39] Shama, D. Dixit, G. Sheet, and Y. Singh, Unusual Magnetotransport from two-dimensional Dirac Fermions in $Pd_3Bi_2Se_2$, *Phys. E: Low-dimens.* **144,**115457 (2022).

[40] Z.-T. Liu, C. Zhang, Q.-Y. Wu, H. Liu, B. Chen *et al*, Charge density wave order and electron-boson coupling in ternary superconductor $Bi_2Rh_3Se_2$, *Sci. China-Phys. Mech. Astron*. **66**, 277411 (2023).

[41] E. Liu, Y. Sun, N. Kumar, L. Muechler, A. Sun, L. Jiao, S. Y. Yang, D. Liu, A. Liang, Q. Xu *et al*, Giant anomalous Hall effect in a ferromagnetic kagome-lattice semimetal, *Nat. phys*, ***14,*** 1125 (2018).

[42] Topological Materials Database https://topologicalquantumchemistry.org/#/, https://www.cryst.ehu.es/

[43] B. Bradlyn, J. Cano, Z. Wang, M. G. Vergniory, C. Felser, R. J. Cava, and B. A. Bernevig, Beyond Dirac and Weyl fermions: Unconventional quasiparticles in conventional crystals, *Science* **353**, aaf5037 (2016).

[44] M. G. Vergniory, L. Elcoro, C. Felser, N. Regnault, B. A. Bernevig and Z. Wang, A complete catalogue of high-quality topological materials, *Nature*, **566** 480 (2019).

[45] B. Bradlyn, L. Elcoro, J. Cano, M. G. Vergniory, Z. Wang, C. Felser, M. I. Aroyo, and B. A. Bernevig, Topological quantum chemistry, *Nature* **547,** 298 (2017).

[46] M. G. Vergniory, B. J. Wieder, L. Elcoro, S. S. Parkin, C. Felser, B. A. Bernevig, and N. Regnault, All topological bands of all nonmagnetic stoichiometric materials, *Science* **376**, 6595 (2022).

[47] Z. Zhang, M. Ikeda, M. Utsumi, Y. Yamamoto, H. Goto, R. Eguchi, Y. F. Liao, H. Ishii, Y. Takabayashi, K. Hayashi, and R. Kondo, Pressure dependence of the structural and superconducting properties of the Bi-based superconductor $Bi_2Pd_3Se_2$, *Inorg. Chem. 63,* 2553 (2024).

[48] R. Weihrich, S. F. Matar, I. Anusca, F. Pielnhofer, P. Peter, F. Bachhuber, and V. Eyert, Palladium site ordering and the occurrence of superconductivity in $Bi_2Pd_3Se_{2-x}S_x$, *J. Solid State Chem.* **184**, 797 (2011).

[49] S. Seidlmayer, F. Bachhuber, I. Anusca, J. Rothballer, M. Bräu, P. Peter, and R. Weihrich, Half antiperovskites: V. Systematics in ordering and group-subgroup-relations for $Pb_2Pd_3Se_2$, $Bi_2Pd_3Se_2$, and $Bi_2Pd_3S_2$, *Z. Kristallogr.- Cryst. Mater.* **225**, 371 (2010)
https://doi.org/10.1524/zkri.2010.1272

[50] Supplemental information includes the XRD and EDX spectrums and structural and ICP-OES analysis (Tables S1-S2, S4), low-temperature transport data, Dingle plot, Landau fan diagrams, extended zone scheme highlighting the open orbits, and details of the ellipsoidal model. Additionally, DFT geometry for the primitive cell (Table S3) and various parameters obtained from dHvA oscillations (Table S4) are included.

[51] J. P. Perdew, A. Ruzsinszky, G. I. Csonka, A. O. Vydrov, G. E. Scuseria, L. A. Constantin, X. Zhou, K. Burke, Restoring the density-gradient expansion for exchange in solids and surfaces, *Phys. Rev. Lett.* **100**, 13 (2008).

[52] P. E. Blöchl, Projector augmented-wave method, *Phys. Rev. B* **50**, 17953 (1994).

[53] A. Kokalj, XCrySDen-a new program for displaying crystalline structures and electron densities, *J. Mol. Graph. Model.* **17**, 176 (1999).

[54] C2x: A tool for visualisation and input preparation for Castep and other electronic structure codes, M. J. Rutter, *Comput. Phys. Commun.* **225** 174 (2018).

[55] A. C. Jacko, J. O. Fjarestadt, B. J. Powell, A unified explanation of the Kadowaki-Woods ratio in strongly correlated metals, *Nat. Phys.* **5**, 423 (2009).

[56] R. Chapai, M. P. Smylie, H. Hebbeker, D. Y. Chung, W.-K. Kwok, J. F. Mitchell, and U. Welp, Superconducting properties and gap structure of the topological superconductor candidate $Ti_3Sb$, *Phys. Rev.* B **107,** 104504 (2023).

[57] T. Baumgartner, M. Eisterer, H. W. Weber, R. Flukiger, C. Scheuerlein and L. Bottura, Effects of neutron irradiation on pinning force scaling in state-of-the-art $Nb_3Sn$ wires, *Supercond. Sci. Technol.* **27**, 015005 (2014).

[58] N. R. Werthamer, E. Helfand, and P. C. Hohenberg, Temperature and purity dependence of the superconducting critical field, $H_{c2}$. III. Electron spin and spin-orbit Effects, *Phys. Rev.* **147**, 295 (1966).

[59] F. Duan and J. Guojun, *Introduction to Condensed Matter Physics*. Vol. 1 (World Scientific, Singapore, 2005).

[60] A. A. Abrikosov, *Fundamentals of the Theory of Metals* (North-Holland, Amsterdam, 1988).

[61] A. B. Pippard, *Magnetoresistance in Metals*, (Cambridge University Press, Cambridge, 1989).

[62] C. M Hurd, *The Hall Effect in Metals and Alloys*, (Plenum Press, New York, 1972).

[63] S. Zhang, Q. Wu, Y. Liu, and O. V. Yazyev, Magnetoresistance from Fermi surface topology, *Phys. Rev. B* **99**, 035142 (2019).

[64] C. Coughlan, M. Ibanez, O. Dobrozhan, A. Singh, A. Cabot, K. M. Ryan, Compound copper chalcogenide nanocrystals, *Chem. Rev*. **117**, 5865 (2017).

[65] D. Shoenberg, *Magnetic Oscillations in Metals* (Cambridge University Press, Cambridge, 2009).

[66] I. E. Lifshitz and A. M. Kosevich, On the theory of magnetic susceptibility of metals at low temperatures, *Sov. Phys. JETP* **2**, 636 (1956).

[67] J. Hu, Z. Tang, J. Liu, X. Liu, Y. Zhu *et al*., Evidence of topological nodal-line fermions in ZrSiSe and ZrSiTe, *Phys. Rev. Lett*. **117**, 016602 92016).

[68] C. Li, C. M. Wang, B. Wan, X. Wan, H. -Z. Lu, X. C. Xie, Rules for phase shifts of quantum oscillations in topological nodal-line semimetals, *Phys. Rev. Lett*. **120**,146602 (2018).

[69] N.W. Ashcroft and N. D. Mermin, *Solid State Physics* (Holt, Rinehart and Wiston, New York, 1976).

[70] R. Chapai, M. Leroux, V. Oliviero, D. Vignolles, N. Bruyant, M. P. Smylie, D. Y. Chung, M. G. Kanatzidis, W.-K. Kwok, J. F. Mitchell, U. Welp, Magnetic breakdown and topology in the Kagome superconductor $CsV_3Sb_5$ under high magnetic field, *Phys. Rev. Lett*. **130**, 126401 (2023).

[71] M. A. Khan, D. E. Graf, I. Vekhter, D. A. Browne, J. F. DiTusa, W. A. Phelan and D. P. Young, Quantum oscillations and a nontrivial Berry phase in the noncentrosymmetric topological superconductor candidate BiPd, *Phys. Rev*. B **99**, 020507 (2019).

[72] R. Chapai, D. A. Browne, D. E. Graf, J. F. DiTusa, and R. Jin, Quantum oscillations with angular dependence in $PdTe_2$ single crystals, *J. Phys.: Condens. Matter* **33**, 035601 (2021).

[73] Z. J. Xiang, D. Zhao, Z. Jin, C. Shang, L K. Ma, G. J. Ye, B. Lei, T. Wu, Z. C. Xia and X. H. Chen, Angular-dependent phase factor of Shubnikov–de Haas oscillations in the Dirac semimetal $Cd_3As_2$, *Phys. Rev. Lett.* **115**, 226401 (2015).

[74] Y. Yang, H. Xing, G. Tang, C. Hua, C. Yao, X. Yan, Y. Lu, J. Hu, Z. Mao, and Y. Liu, Anisotropic Berry phase in the Dirac nodal-line semimetal ZrSiS: The effect of spin-orbit coupling, *Phys. Rev. B* **103**, 125160 (2021).

[75] J. Cao, S. Liang, C. Zhang, Y. Liu, J. Huang, Z. Jin, Z. G. Chen, Z. Wang, Q. Wang, J. Zhao and S. Li, Landau level splitting in $Cd_3As_2$ under high magnetic fields, *Nat. Commun.* **6**, 7779 (2015).

[76] R. Chapai, P. S. Reddy, L. Xing, D. E. Graf, A. B. Karki, T.-R. Chang, and R. Jin, Evidence for unconventional superconductivity and nontrivial topology in PdTe, *Sci. Rep.* **13** 6824 (2023).

[77] A. Alexandradinata and L. Glazman, Fermiology of topological metals, *Annu. Rev. Condens. Matter Phys*. **14**, 261 (2023).

[78] K. W. Chen, G. Zheng, D. Zhang, A. Chan, Y. Zhu, K. Jenkins, F. Yu, M. Shi, J. Ying, Z. Xiang, X. Chen, Z. Wang, J. Singleton, and L. Li, 2023. Magnetic breakdown and spin-zero effect in quantum oscillations in kagome metal $CsV_3Sb_5$, *Commun. Mater*. **4**, 96(2023).

[79] P. Li, Y. Wen, X. He, Q. Zhang, C. Xia, Z. M. Yu, S. A. Yang, Z. Zhu, H. N. Alshareef, and X. X. Zhang, Evidence for topological type-II Weyl semimetal WTe$_2$, *Nat. commun.* **8**, 2150 (2017).

## SUPPLEMENTAL MATERIALS

**Crystal structure and structural refinement**

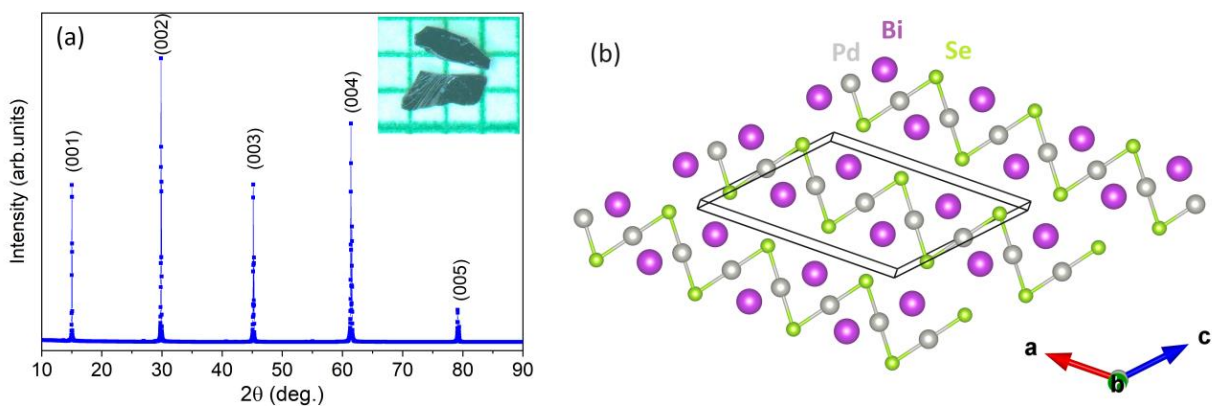

FIG. S1. (a) XRD pattern of an oriented single crystal of Pd$_3$Bi$_2$Se$_2$ indexed in the monoclinic structure with space group C2/*m* (#12). The indexed peaks correspond to the (00*l*) plane. Inset: Optical images of Pd$_3$Bi$_2$Se$_2$ single crystals placed on a millimeter grid. (b) Schematic crystal structure of monoclinic Pd$_3$Bi$_2$Se$_2$. A pseudo-2D view is shown, with Bi-Se and Bi-Pd bonds omitted for clarity.

Table S1. Crystal data and structure refinement for Pd$_3$Bi$_2$Se$_2$.
Numbers in parentheses are estimated standard deviations

| Empirical formula | Pd$_3$Bi$_2$Se$_2$ |
|---|---|
| Formula weight | 895.08 |
| Temperature/K | 296 |
| Crystal system | monoclinic |
| Space group | C2/*m* |
| *a*/Å | 11.7403(8) |
| *b*/Å | 8.4339(6) |
| *c*/Å | 8.4163(6) |
| *α*/° | 90 |
| *β*/° | 133.834(1) |
| *γ*/° | 90 |
| Volume/Å$^3$ | 601.13(7) |
| Z | 4 |
| $\rho_{calc}$, g/cm$^3$ | 9.890 |
| *μ*/mm$^{-1}$ | 79.117 |
| *F*(000) | 1488.0 |
| Crystal size/mm$^3$ | 0.101 × 0.093 × 0.086 |
| Radiation | MoKα (λ = 0.71073 Å) |

| 2Θ range for data collection/° | 6.712 to 52.688 |
| --- | --- |
| Index ranges | $-14 \leq h \leq 14, -10 \leq k \leq 10, -10 \leq l \leq 10$ |
| Reflections collected | 10110 |
| Completeness | 100% |
| Independent reflections | 661 [$R_{int}$ = 0.0597, $R_{sigma}$ = 0.0224] |
| Data/restraints/parameters | 661/0/40 |
| Goodness-of-fit on $F^2$ | 1.197 |
| Final $R$ indexes [$I>=2\sigma (I)$] | $R_1$ = 0.0207, $wR_2$ = 0.0577 |
| Final $R$ indexes [all data] | $R_1$ = 0.0210, $wR_2$ = 0.0578 |
| Largest diff. peak/hole / e Å$^{-3}$ | 3.07/-1.75 |

Table S2. Atomic coordinates ($\times 10^4$) and equivalent isotropic displacement parameters (Å$^2 \times 10^3$) for Pd$_3$Bi$_2$Se$_2$. $U_{eq}$ is defined as 1/3 of of the trace of the orthogonalised $U_{ij}$ tensor. Numbers in parentheses are estimated standard deviations

| Atom | Wyckoff position | x | y | z | U(eq) | Occupancy |
| --- | --- | --- | --- | --- | --- | --- |
| Bi(1) | 4i | 2186.0(6) | 0 | 7028.8(9) | 9.26(18) | 1 |
| Bi(2) | 4i | 2499.7(6) | 0 | 12586.7(8) | 6.95(18) | 1 |
| Pd(1) | 4i | 265.9(12) | 0 | 8015.4(18) | 9.3(3) | 1 |
| Pd(2) | 4f | 2500 | 2500 | 5000 | 7.5(3) | 1 |
| Pd(3) | 4h | 0 | -2768.0(13) | 5000 | 8.7(3) | 1 |
| Se(1) | 8j | 336.7(11) | 2901.7(12) | 8217.2(16) | 8.1(2) | 1 |

Table S3. DFT geometry for the primitive cell

| | a (Å) | b (Å) | c (Å) |
| --- | --- | --- | --- |
| | 7.21061 | 7.21061 | 8.41904 |
| | | | |
| | α (°) | β (°) | γ (°) |
| | 55.9731 | 124.0269 | 108.2586 |
| | | | |
| Bi1 | 0.21825 | 0.78175 | 0.70262 |
| Bi2 | 0.78175 | 0.21825 | 0.29738 |
| Bi3 | 0.25017 | 0.74983 | 0.25896 |
| Bi4 | 0.74983 | 0.25017 | 0.74104 |
| Pd1 | 0.02627 | 0.97373 | 0.80249 |
| Pd2 | 0.97373 | 0.02627 | 0.19751 |
| Pd3 | 0.50000 | 0.00000 | 0.50000 |
| Pd4 | 0.00000 | 0.50000 | 0.50000 |
| Pd5 | 0.72361 | 0.72361 | 0.50000 |
| Pd6 | 0.27639 | 0.27639 | 0.50000 |
| Se1 | 0.32317 | 0.25626 | 0.82134 |
| Se2 | 0.25626 | 0.32317 | 0.17866 |
| Se3 | 0.67683 | 0.74374 | 0.17866 |
| Se4 | 0.74374 | 0.67683 | 0.82134 |

**Energy dispersive X-ray (EDX) spectroscopy**

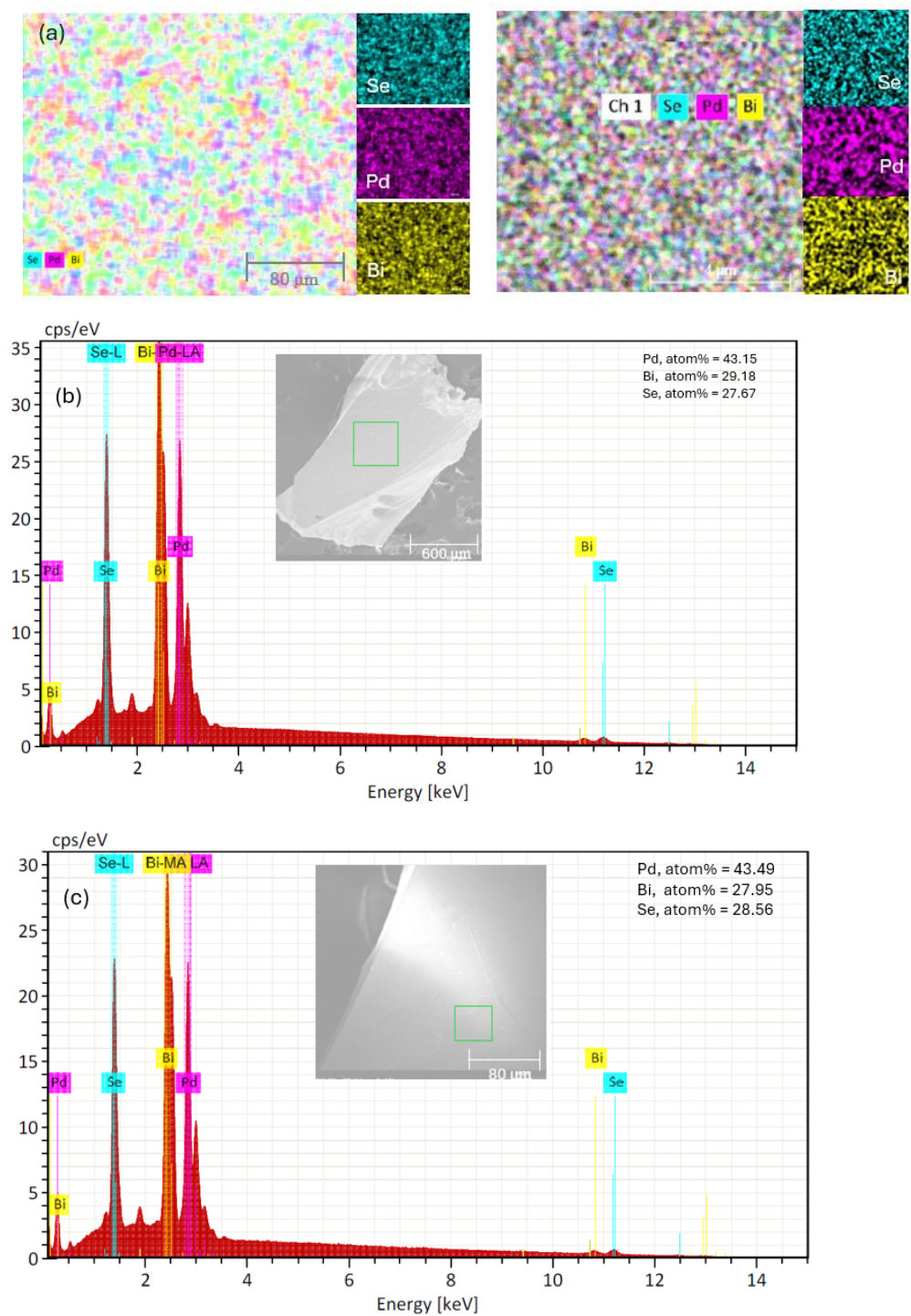

FIG. S2. (a) Energy dispersive X-ray spectroscopy (EDX) mapping of elemental composition in $Pd_3Bi_2Se_2$ single crystal. (b-c) EDX spectrum of $Pd_3Bi_2Se_2$.

Multiple spots with different areas in the samples were examined. The EDX analysis of the composition yields an average atomic percentage of (43.32 ± 0.17) % (Pd): (28.57 ±0.61) % (Bi):

(28.12 ± 0.45) % (Se), in excellent agreement with 42.8%: 28.6%: 28.6% expected for stoichiometric $Pd_3Bi_2Se_2$.

**Inductively coupled plasma-optical emission spectroscopy (ICP-OES)**

Several single crystals of $Pd_3Bi_2Se_2$ with total mass of approximately 58 milligrams was dissolved in $HNO_3$ and HCl. The dissolved sample was diluted to 50 mL for the ICP-OES measurement. ICP-OES analysis yields an average weight percent of (33.85 ± 1.70) % (Pd): (46.3 ± 2.30) % (Bi): (19.45 ± 0.97) % (Se) in good agreement with 35.7%: 46.7%:17.6% expected for stoichiometric $Pd_3Bi_2Se_2$.

Table S4: Chemical composition of $Pd_3Bi_2Se_2$ measured by Inductively Coupled Plasma-Optical Emission Spectroscopy. The results are presented in weight percent with an uncertainty of ±5%.

| Sample wt. (g) | Pd, weight % | Bi, weight% | Se, weight% |
|---|---|---|---|
| 0.03020 | 33.9 ± 1.70 | 46.4 ± 2.30 | 19.4 ±0.97 |
| 0.02770 | 33.8 ± 1.70 | 46.2 ± 2.30 | 19.5 ±0.97 |

**Low temperature magneto-transport**

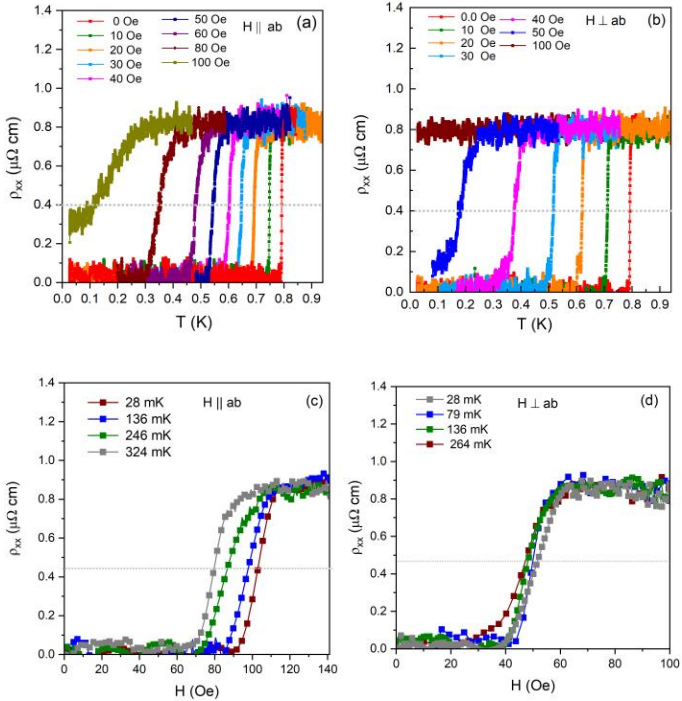

FIG. S3 Electrical resistivity of $Pd_3Bi_2Se_2$ with temperature sweep at fixed field (a-b), with field sweep at fixed temperature (c-d).

These data are used to construct the phase diagram in Fig. 2 in the main text. In transport data (Fig. S3), we have used 50% criterion as indicated by the horizontal dotted lines in Fig. S3 to estimate $T_c$ and $H_{c2}$, while in TDO data (Fig. 1(b-c)), as the 50% criteria is not fully satisfied by all curves, we implemented the 10% criterion.

**Kohler's scaling rule**

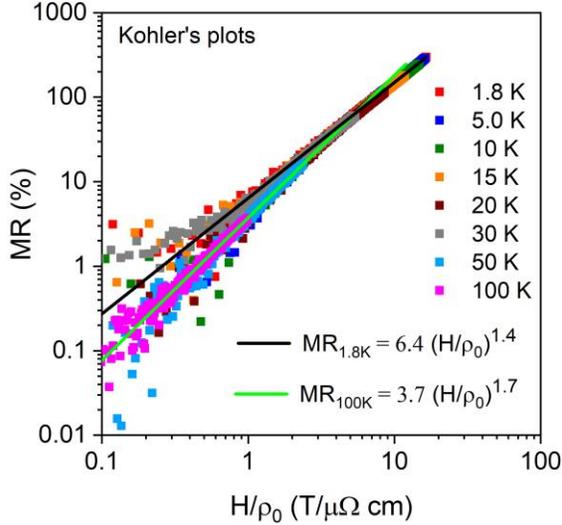

FIG. S4. Kohler scaling plots of $Pd_3Bi_2Se_2$. Solid lines fit with the power law to the data at 1.8 K (black) and 100 K(green). The slope of MR vs. $H$ in high fields is seen to decrease from 1.7 at low $H/\rho(0)$ to 1.4.

**FFT spectrum highlighting higher frequencies**

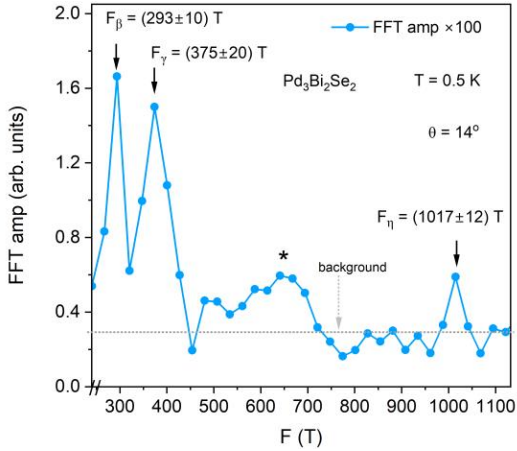

FIG. S5. FFT amplitude of as a function of frequency at T = 0.5 K and $\theta = 14°$. Frequencies in a window of 240 T to 1120 T are displayed highlighting $F_\beta$, $F_\gamma$ and $F_\eta$. Here, the amplitude is multiplied by 100 as compared to the spectrum in Fig.4 in the main text. The error bars are taken as the half-width at the half-height of the FFT peaks. There is a broad peak around 640 T, denoted by an asterisk (*), which is difficult to assign to a specific frequency and is left unidentified.

## Analysis of quantum oscillations and Dingle plot

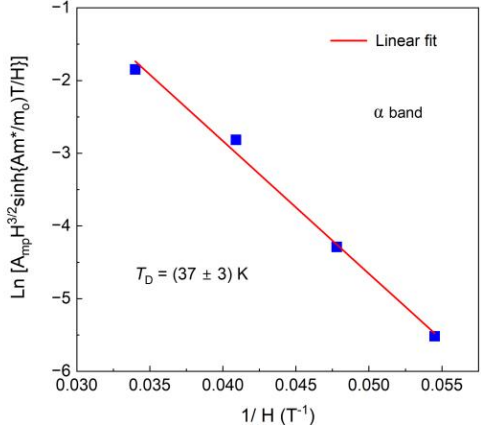

FIG. S6. Dingle plot for the α band of $Pd_3Bi_2Se_2$ constructed from dHvA oscillations data presented in Fig. 4(a) in the main text.

Table S5: Parameters obtained from the dHvA oscillations in $Pd_3Bi_2Se_2$ including the oscillation frequency ($F$), the Fermi wave vector ($k_F$), the effective mass ($m^*$), the Fermi velocity ($v_F$), the Dingle temperature ($T_D$), the quantum relaxation time ($\tau_q$), the quantum mobility ($\mu_q$), and the mean free path ($\lambda_q$).

| Band | F(T) | $S_F(10^{-2}\text{Å}^{-2})$ | $k_F(\text{Å}^{-1})$ | $m^*/m_0$ | $v_F(10^5 \text{ m/s})$ | $T_D$(K) | $\tau_q(10^{-13}\text{ s})$ | $\mu_q(\text{m}^2\text{V}^{-1}\text{s}^{-1})$ | $\lambda_q$ (nm) |
|---|---|---|---|---|---|---|---|---|---|
| α | 150 | 1.430 | 0.067 | 0.11 | 6.99 | 37±3 | 0.32 | 0.050 | 22.3 |
| β | 293 | 2.793 | 0.094 | – | – | – | – | – | – |
| γ | 375 | 3.575 | 0.106 | – | – | – | – | – | – |
| η | 1017 | 9.697 | 0.175 | – | – | – | – | – | – |

## Angular dependence of magnetic torque

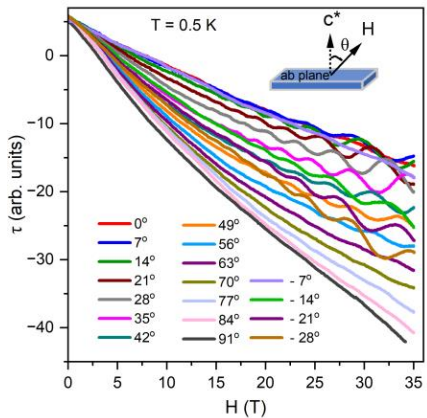

FIG. S7. Magnetic torque, $\tau(H)$, of $Pd_3Bi_2Se_2$ measured for different angles $\theta$ (defined in the inset) at $T = 0.5$ K. These data are presented in Fig. 5(a) after smooth background (third order polynomial) subtraction.

As mentioned in the main text, the magnetic torque, τ(H), of $Pd_3Bi_2Se_2$ was measured for different angles at various temperatures. Figure S7 shows the raw data of magnetic torque, τ(H), measured for different angles θ (defined in the inset) at T = 0.5 K. With increasing θ, there is a systematic increase of the torque signal for positive and negative angles as we rotate the field with respect to the *ab*-plane of the crystal. In our experiment, with a 7-degree angular interval, 14 degrees was selected for temperature and field dependent measurement because it is close to the angle where the field is nearly perpendicular to the ab plane, which enhances the oscillations. At higher angles, the oscillations begin to dampen due to reduced electron mobility in the inter-layer direction, leading to an increased Dingle temperature and suppression of the oscillations. This behavior is detailed in the angular dependence data presented in Fig. 5 in the main text. Although a zero-degree angle would theoretically optimize electron mobility, the torque signal at this angle is too weak to be reliably detected across various temperatures for estimating effective mass.

**Landau level fan diagrams**

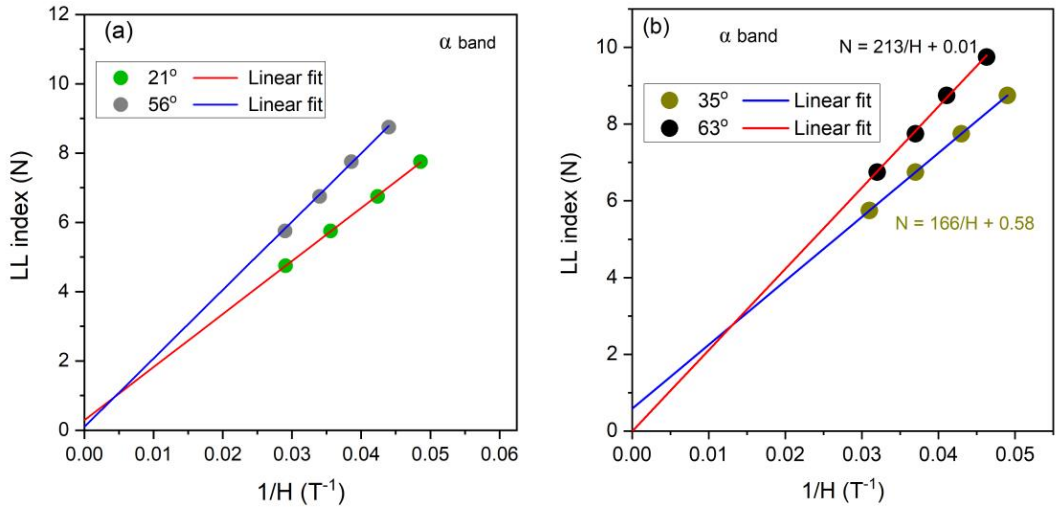

FIG. S8 (a-b). Landau level fan diagrams constructed from the dHvA data presented in Fig. 5(a) (main text) for the α band at selected angles as indicated.

The changing intercepts in these fan diagrams clearly indicate the angular dependence of the phase factor in the torque data. We note though that, this change itself cannot be considered as arising solely due to Berry phase change, as there are additional phase factors, including orbital moment and Zeeman coupling, contributing to the phase shift in dHvA oscillations [77].

**Extended zone scheme highlighting the open orbits**

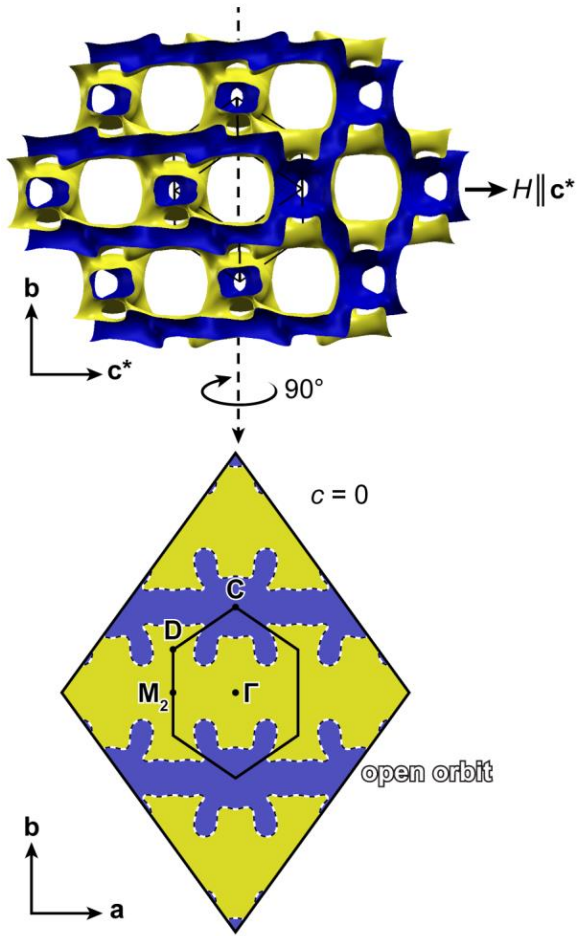

FIG. S9. The Fermi surface derived from band #145 contains tubular structures oriented parallel to the ab-planes. These carry open electron orbits for magnetic fields applied along the c*-direction as highlighted in the lower panel and may thus account for the magneto-transport behavior of $Pd_3Bi_2Se_2$.

**Ellipsoidal model**

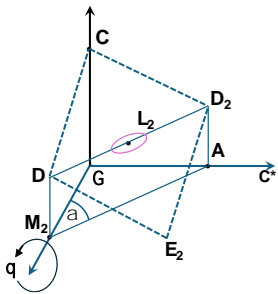

FIG. S10. Geometry used in the analysis of the angular dependence of quantum oscillations. The location of the $L_2$-pocket and various high-symmetry points in the Brillouin zone (BZ) are shown.

Our Pd$_3$Bi$_2$Se$_2$ single crystals grow in the shape of plates in which the large faces are parallel to the internal layering of the crystal structure. In our experiments, the normal to the layers, c*, corresponds to the zero-angle magnetic field direction. We model the L$_2$-pocket as an ellipsoid of revolution centered on the D-D$_2$ line in the BZ as shown in the schematic. Thus, the long axis of the ellipsoid suspends an angle of $\alpha \sim 45°$ with c*. One also notes that the D-D$_2$ line is parallel to the A-M$_2$ line. Then, the geometry shown in Fig. S10 yields the following relation for the angular dependence of the dHvA frequency for rotations by angle $\theta$ around the $\Gamma$-M$_2$ axis which was used to fit the data in Fig. 5(c).

$$F(\theta) = \frac{2F_{min}}{\sqrt{4AB - C^2}}$$

where
$A = \cos^2\theta + \sin^2\theta\,(\cos^2\alpha + \Gamma^2 \sin^2\alpha)$
$B = \sin^2\alpha + \Gamma^2 \cos^2\alpha$
$C = 2\,(1 - \Gamma^2)\,\sin\theta\,\cos\alpha\,\sin\alpha$

and $\Gamma = F_{min}/F_{max} \leq 1$

Here, $F_{min}$ corresponds to the circular equatorial cross section perpendicular to D-D$_2$ while the minimum measured frequency, $F(\theta = 0)$, is given as $F(\theta = 0) = F_{min}/\sqrt{\sin^2\alpha + \Gamma^2 \cos^2\alpha}$. $F_{max}$ corresponds to any cross section containing the D-D$_2$ line, i.e., $F_{max} = F(\theta = \pi/2) = F_{min}/\Gamma$.